\documentclass[a4paper, 12pt]{article}
\usepackage[dvips]{graphicx}

\def\ltap{\ \raise.3ex\hbox{$<$\kern-.75em\lower1ex\hbox{$\sim$}}\ }
\def\gtap{\ \raise.3ex\hbox{$>$\kern-.75em\lower1ex\hbox{$\sim$}}\ }
\def\gl{\ \raise.4ex\hbox{$>$\kern-.75em\lower1ex\hbox{$<$}}\ }

\begin{document}

\title{Triquark structure and isospin symmetry breaking in exotic $D_{s}$ mesons}
\author{S.~Yasui and M.~Oka \\
\normalsize Department of Physics, Tokyo Institute of Technology,\\
\normalsize Tokyo 152-8551, Japan \\}
\maketitle

\begin{abstract}
The color anti-triplet triquark $q\bar{q}\bar{q}$ is considered as a compact component in the tetraquark structure $cq\bar{q}\bar{q}$ of exotic $D_{s}$ mesons.
We discuss the mass spectrum and the flavor mixing of the triquarks by using the instanton induced interaction and the one-gluon exchange potentials.
As a characteristic property of the triquark, we investigate the isospin violation.
It is shown that the flavor $\overline{\bf 3}$ (isosinglet) and $\bf 6$ (isotriplet) states may be strongly mixed and then are identified with $D_{s}(2632)$.
\end{abstract}

\section{Introduction}

Exotic $D_{s}$ mesons attract much attention recently.
The BaBar Collaboration \cite{BaBar_03} first announced the $D_{s}(2317)$ with $J^{\pi}=0^{+}$, whose mass lies approximately 160 MeV below the constituent quark model predictions \cite{Godfrey_Isgur_85, Cahn_Jackson_03}.
The decay width is less than $4.6$ MeV.
This state was confirmed by CLEO \cite{CLEO_03} and Belle \cite{Belle_03}.
It was followed by $D_{s}(2460)$ with $1^{+}$, reported by CLEO \cite{CLEO_03}, and $D_{s}(2632)$ by SELEX \cite{SELEX_04}.
At the same time, a charmonium candidate $X(3872)$ was reported by Belle \cite{Belle_03a} and the other groups.
These new mesons are novel because they do not fit to the quark model expectations and have caused further studies and speculations on their structures.
In particular, it has been proposed that some of them are excited two-quark states \cite{Beveren_Rup_03, Matsuki_etal_07},
chiral doublets in heavy quark limit \cite{Bardeen_etal_03},
the molecular bound states \cite{Barnes_etal_03, Szczepaniak_03}
or tetraquarks $cq\bar{q}\bar{q}$ ($q=u$, $d$ and $s$) \cite{Cheng_Hou_03, Chen_Li_04, Swanson_06,Hayashigaki_Terasaki_06, Maiani_etal_04, Maiani_etal_05, Liu_Zhu_04}.

Here we briefly review the previous researches about the molecular and tetraquark pictures.
Some suggest that the $D_{s}(2317)$ may be a $DK$ molecule state, and the $D_{s}(2460)$ a $D^{*}K$ molecule.
Indeed the masses of the $D_{s}(2317)$ and $D_{s}(2460)$ are slightly below the thresholds of the $D+K$ and $D^{*}+K$, respectively.
The mass splitting 140 MeV between $D_{s}$ and $D_{s}^{\ast}$ is almost the same as that  between the $D_{s}(2317)$ and $D_{s}(2460)$.
The ground state of $D_{s}$ is the $D_{s}(1969)$ with $0^{-}$ and the $D_{s}(2112)$ with $1^{-}$.
Therefore, the $D_{s}(2317)$ can be assigned to be the ground state in the $0^{+}$ sector, while the $D_{s}(2460)$ can be identified also as a ground state in the $1^{+}$ sector.
These properties are also explained in the chiral effective theory \cite{Bardeen_etal_03}.

One of the prominent properties of the exotic $D_{s}$ mesons is the isospin violating decay process $D_{s}(2317) \rightarrow D_{s}\pi^{0}$ \cite{BaBar_03}.
This process is considered to be realized by the virtual $\eta$ emission and the $\eta-\pi$ mixing, 
since the $D_{s}(2317)$ is supposedly an isosinglet state.\footnote{Hayashigaki and Terasaki considers a possibility of the isotriplet state for $D_{s}(2317)$ \cite{Hayashigaki_Terasaki_06}.}
An anomalous branching ratio $\Gamma(D_{s}(2632) \rightarrow D^{0}K^{+})/\Gamma(D_{s}(2632) \rightarrow D_{s}\eta) \simeq 0.16 \pm 0.06$ \cite{SELEX_04} is also a very interesting problem.
In the conventional $c\bar{s}$ picture, the decay to the $D^{0}K^{+}$ is favored as compared with the decay to $D_{s}\eta$, since $u\bar{u}$ or $d\bar{d}$ creation would be easier than $s\bar{s}$.
Maiani {\it et al.} \cite{Maiani_etal_04} considered the tetraquark state $c\bar{s}d\bar{d}$, in which the isospin is maximally violated.
This state has decay modes of $[c\bar{s}][d\bar{d}]$ ($D_{s}\eta$) and $[c\bar{d}][d\bar{s}]$ ($D^{+}K^{0}$), 
while the $D^{0}K^{+}$ decay is suppressed by the OZI forbidden process $d\bar{d} \rightarrow u\bar{u}$.
This picture was also applied to study of the isospin violation in the decay process of $X(3872)$ \cite{Maiani_etal_05}.
On the other hand, Chen and Li considered $c\bar{s}s\bar{s}$ \cite{Chen_Li_04}.
They discussed that the decay to $D_{s}\eta$ is dominant, while the decay to $D^{0}K^{+}$ is suppressed by $1/N_{c}$ due to the OZI rule and the reduction of the amplitude due to the color matching to create a color singlet state.
Liu and Zhu discussed that the $D_{s}(2632)$ is assigned as an isosinglet member in flavor $\overline{ \bf{15}}$ \cite{Liu_Zhu_04}.

Let us see the possibility of the isospin violated eigenstates \cite{Maiani_etal_04, Maiani_etal_05}.
When the quarks have sufficiently large momenta, the asymptotic freedom suppresses the $q\bar{q}$ creation process, and the flavor mixing interaction is less important.
Then, the alignment in the diagonal components in the mass matrix realizes $u\bar{u}$ and $d\bar{d}$ separately as eigenstates, which are mixed states of isosinglet and isotriplet states.
In general, however, the flavor mixing term is in the order of $\sim$100 MeV \cite{Oka_Takeuchi_89, Oka_Takeuchi_91, Takeuchi_94, Takeuchi_96, Shinozaki_etal_05}, and much larger than  the mass difference between $u$ and $d$ quarks, $|m_{u}-m_{d}| \ltap 5$ MeV.
Therefore, it is expected that the isospin breaking effect is too small to separate $u\bar{u}$ and $d\bar{d}$.

The purpose of this paper is to discuss the microscopic mechanism of the isospin violation of the tetraquark for the open charm system, $cq\bar{q}\bar{q}$.
It is noticed that the interaction between the light quarks $q$ and the $c$ quark is suppressed in the heavy quark limit.
Thus, it is natural to consider that, in the first approximation, three light quarks are decoupled from the heavy quark.
Therefore we consider states compound by the $u$, $d$  and $s$ quarks as triquarks or color non-singlet baryons.
Here it must be noticed such a state cannot exist as an asymptotic state, but only in the bound state. 

We here consider a simple model with non-relativistic valence quarks under the influence of the one-gluon exchange (OGE) and the instanton induced interaction (III).
The mass spectroscopy of the triquark was first discussed in the diquark-triquark picture in the literature of the pentaquark \cite{Karliner_Lipkin_03, Kochelev_etal_04, Lee_etal_05}, and further investigated in details in the OGE interaction \cite{Hogaasen_Sorba_04, Delgado_05, Jaffe_05}.
Furthermore, the 't Hooft interaction induced by the instanton was also used \cite{Dmitra_04, Dmitra_05, Dmitra_06}.
However, the effective interaction employed in \cite{Dmitra_04, Dmitra_05, Dmitra_06} operates only in spin singlet and isosinglet channel in $q\bar{q}$ pair, 
while the effective interaction used in \cite{Oka_Takeuchi_89, Oka_Takeuchi_91, Takeuchi_94, Takeuchi_96, Shinozaki_etal_05} operates, not only in spin singlet and isosinglet channel, but also in spin triplet and isotriplet channel. 
It is known that the difference causes a discrepancy in the meson mass spectrum \cite{Dmitra_05a}.
Therefore it is an interesting problem to investigate the isospin violation of the triquark by using the effective interaction in  \cite{Oka_Takeuchi_89, Oka_Takeuchi_91, Takeuchi_94, Takeuchi_96, Shinozaki_etal_05}.

The content of this paper is as follows.
In Section 2, the flavor representation of the triquark, the III and OGE potential and the mass matrix are discussed.
In Section 3, the isospin mixing  is investigated by considering the $ud$ quark mass difference.
In Section 4, our discussion is summarized.

\section{Quark model}

\begin{figure}[tbp]
\begin{center}
\includegraphics[width=7cm, angle=0, clip]{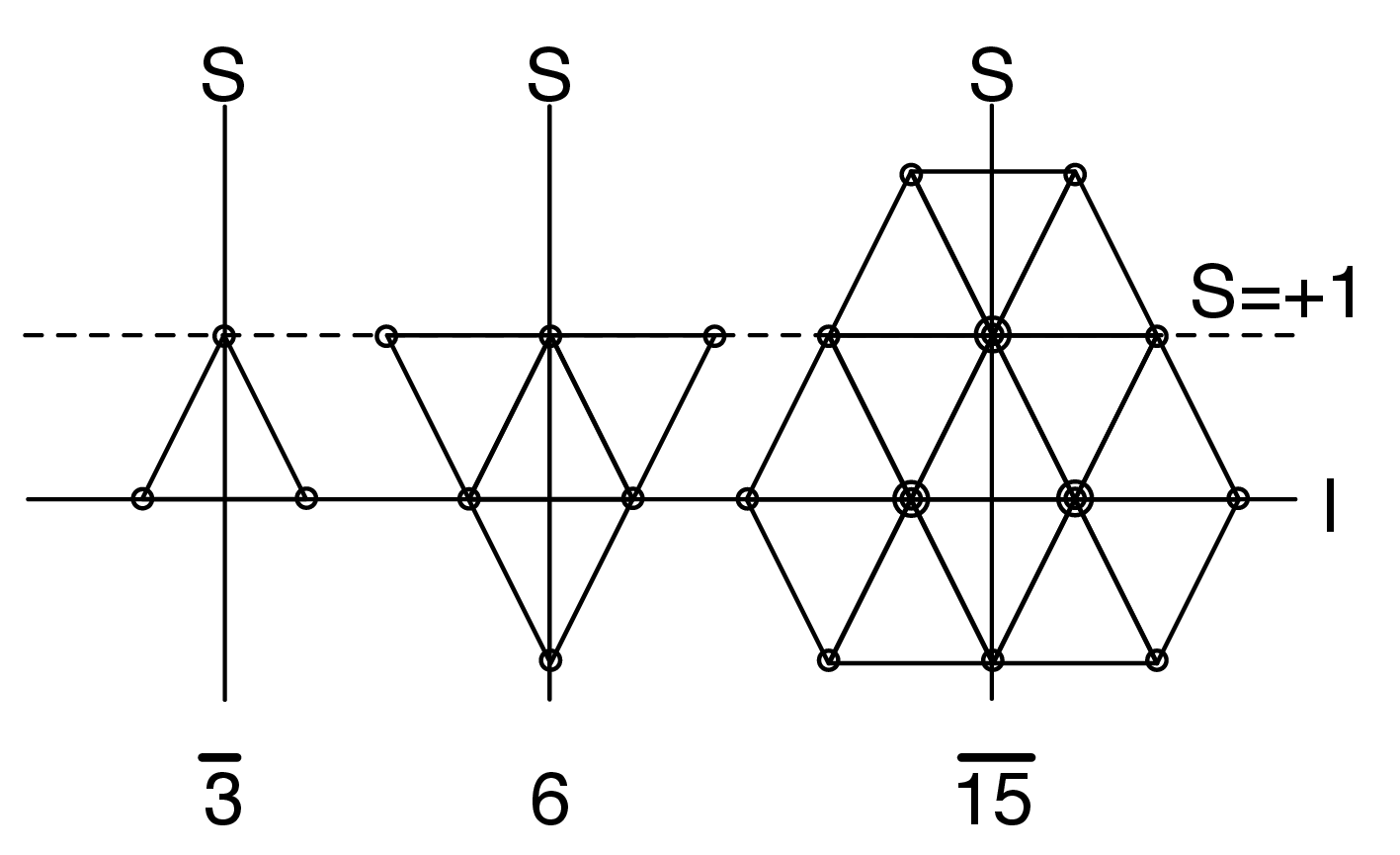}
\end{center}
\vspace*{0.0cm} \caption{\small \baselineskip=0.5cm The weight diagram of the flavor $SU(3)$ multiplets of the triquark $q\bar{q}\bar{q}$.}
 \label{fig : multiplets}
\end{figure}

In the tetraquark picture of the exotic $D_{s}$ mesons, the triquark is considered as a bound state composed by three light flavor quarks.
The hamiltonian of the triquark is obtained only in light flavors space, since the interaction between the light and heavy quarks is suppressed in the OGE potential.
This is also the case for the instanton induced interaction, since the heavy quark has no zero mode and free from the instanton vacuum \cite{'t_Hooft_76, Shifman_etal_80}.
The flavor $SU(3)$ multiplets of the triquark state $q\bar{q}\bar{q}$ are given as
\begin{eqnarray*}
 {\bf 3} \otimes \overline{\bf 3} \otimes \overline{\bf 3} = \overline{\bf 3}_{S} \oplus  \overline{\bf 3}_{A} \oplus {\bf 6}_{A} \oplus \overline{\bf 15}_{S}.
\end{eqnarray*}
We write the subscripts of $S$ and $A$ according to the symmetry under the exchange of two anti-quarks.
In Fig.~\ref{fig : multiplets}, we show the weight diagrams of these multiplets.
It is assumed that all the quarks and anti-quarks occupy the lowest energy single particle orbital, the $s$-wave orbital.
In the following, we omit the subscripts in ${\bf 6}_{A}$ and $\overline{\bf 15}_{S}$ for simplicity.

The exotic states reported in experiments have the strangeness $S=+1$.
Then, the isospin for each flavor multiplet is as follows; isosinglet for $\overline{\bf 3}_{A}$, $\overline{\bf 3}_{S}$ and ${\overline{\bf 15}}^{0}$, and isotriplet for $\bf 6$ and ${\overline{\bf 15}}^{1}$.
Here the isospin components of ${\overline{\bf 15}}$ are distinguished by the superscript.
It is straightforward to write down the flavor wavefunctions of these multiplets for $S=+1$.
\begin{eqnarray}
\begin{array}{l}
\mbox{isosinglet} \left\{
{\renewcommand\arraystretch{2.0}
\begin{array}{l}
| {\overline{\bf 3}_{A}} \rangle = \frac{1}{2} \left[ u(\bar{s}\bar{u}-\bar{u}\bar{s}) - d(\bar{d}\bar{s}-\bar{s}\bar{d}) \right]
 \\ 
| {\overline{\bf 3}_{S}} \rangle = \frac{1}{2\sqrt{2}} \left[ 2s\bar{s}\bar{s} + u(\bar{s}\bar{u}+\bar{u}\bar{s}) + d(\bar{d}\bar{s}+\bar{s}\bar{d}) \right]
   \\
| {\overline{\bf 15}}^{0} \rangle =\frac{1}{2\sqrt{2}} \left[ 2s\bar{s}\bar{s} - u(\bar{s}\bar{u}+\bar{u}\bar{s}) - d(\bar{d}\bar{s}+\bar{s}\bar{d}) \right]
\end{array}
}
\right.  \\
\mbox{isotriplet} \left\{
{\renewcommand\arraystretch{2.0}
\begin{array}{l}
| {\bf 6} \rangle = \frac{1}{2} \left[ u(\bar{s}\bar{u}-\bar{u}\bar{s}) + d(\bar{d}\bar{s}-\bar{s}\bar{d}) \right] \\
| {\overline{\bf 15}}^{1} \rangle = \frac{1}{2} \left[ u(\bar{s}\bar{u}+\bar{u}\bar{s}) - d(\bar{d}\bar{s}+\bar{s}\bar{d}) \right].  
\end{array}
}
\right.
\end{array}
   \label{eq:wavefunction}
\end{eqnarray}
In the following discussion, we consider only the $S=+1$ sector.

The triquark must belong to the color anti-triplet state, ${\overline{\bf 3}}_{S}^{c}$ and ${\overline{\bf 3}}_{A}^{c}$, so that the tetraquark is a color singlet state.
Then, the spin and color combination of the triquark is restricted by the Pauli principle.
For example, the spin and color basis for the flavor ${\overline{\bf 3}}_{A}$ and $\bf 6$ states with the spin $J\!=\!1/2$  is $\{ | \lambda X \rangle, | \rho Y \rangle \}$.
Here, $\lambda$ and $\rho$ stands for the mixed states with $\lambda$- and $\rho$-symmetry in spin $1/2$, and $X$ and $Y$ for color ${\overline{\bf 3}}_{S}^{c}$ and ${\overline{\bf 3}}_{A}^{c}$, respectively.
On the other hand, the basis for the flavor ${\overline{\bf 3}}_{S}$ and $\overline{\bf 15}$ states with spin 1/2 is $\{ | \rho X \rangle, | \lambda Y \rangle \}$.
For spin $J\!=\!3/2$ state, we have $| J\!=\!3/2 \; X \rangle$ for ${\overline{\bf 3}}_{A}$ and $\bf 6$, and $| J=\!3/2 \; Y \rangle$ for ${\overline{\bf 3}}_{S}$ and $\overline{\bf 15}$.

Now we discuss the hamiltonian of the triquark.
The instanton induced interaction (III) has played very important role in the QCD vacuum in accompany with dynamical chiral symmetry breaking \cite{Oka_Takeuchi_89, Oka_Takeuchi_91, Takeuchi_94, Takeuchi_96, Shinozaki_etal_05}.
It induces the Kobayashi-Kondo-Maskawa-'t Hooft (KKMT) interaction \cite{Kobayashi_etal_71, 't_Hooft_76, Shifman_etal_80}, which is given as $2N_{f}$ point-like vertex with the flavor anti-symmetric channel.
In the quark model, the instanton effect has been discussed in the non-relativistic limit in the KKMT interaction.
The OGE potential is also often used as an effective interaction \cite{Rujula_Georgi_Glashow_75}.
Here we consider a hybrid model of the III and OGE potentials \cite{Oka_Takeuchi_89, Oka_Takeuchi_91, Takeuchi_94, Takeuchi_96, Shinozaki_etal_05}.
The hamiltonian is
\begin{eqnarray}
H = K + p_{III} \left[ H_{III}^{(3)} + H_{III}^{(2)} \right] + (1-p_{III}) V_{OGE}  + M_{mass} + V_{conf},
\end{eqnarray}
with the kinetic term $K$, the instanton induced interaction $H_{III}^{(i)}$ ($i=2$ and $3$ for the two- and three-body interactions), the OGE potential $V_{OGE}$, the mass matrix $ M_{mass}$ and the confinement potential $V_{conf}$.
The parameter $p_{III}$ controls the ratio of the III and the OGE potentials.
In the present discussion, we are interested in the isospin symmetry breaking, and not involved with the absolute masses of the tetraquarks.
Therefore, we pick up only the III and OGE terms and the mass matrix;
\begin{eqnarray}
\widetilde{H} = p_{III} \left[ H_{III}^{(3)} + H_{III}^{(2)} \right] + (1-p_{III}) V_{OGE}  + M_{mass}.
\label{eq:hamiltonian2}
\end{eqnarray}
Since we do not solve the quark confinement dynamically, we just use a quark wave function from the harmonic oscillator potential with frequency $\omega$.

Concerning the III potential, the three body force in the three quark state, $q_{1}q_{2}q_{3}$, is given in the flavor diagonal form as
\begin{eqnarray*}
 H_{I\!I\!I}^{(3)} &=& \frac{V_{0}^{(3)}}{4}
 \left[
   1 + \frac{3}{32} \left( \vec{\lambda}_{1} \!\cdot\! \vec{\lambda}_{2} 
                                 + \vec{\lambda}_{2} \!\cdot\! \vec{\lambda}_{3}
                                 + \vec{\lambda}_{3} \!\cdot\! \vec{\lambda}_{1}
                                 \right)
     - \frac{9}{320} d_{abc} \lambda_{1}^{a} \lambda_{2}^{b} \lambda_{3}^{c} \right.
    \nonumber \\
    && \mbox{ } + \frac{9}{32} \left( \vec{\sigma}_{1} \!\cdot\! \vec{\sigma}_{2}  \vec{\lambda}_{1} \!\cdot\! \vec{\lambda}_{2} 
                                 + \vec{\sigma}_{2} \!\cdot\! \vec{\sigma}_{3} \vec{\lambda}_{2} \!\cdot\! \vec{\lambda}_{3}
                                 + \vec{\sigma}_{3} \!\cdot\! \vec{\sigma}_{1} \vec{\lambda}_{3} \!\cdot\! \vec{\lambda}_{1}
                                 \right)
     \nonumber \\
     && \mbox{ } + \frac{27}{320} d_{abc} \lambda_{1}^{a} \lambda_{2}^{b} \lambda_{3}^{c} 
                     \left( \vec{\sigma}_{1} \!\cdot\! \vec{\sigma}_{2} 
                                 + \vec{\sigma}_{2} \!\cdot\! \vec{\sigma}_{3}
                                 + \vec{\sigma}_{3} \!\cdot\! \vec{\sigma}_{1}
                           \right)
     \nonumber \\
     && \mbox{ } - \left.  \frac{9}{64} \epsilon_{ijk} \sigma_{1}^{i} \sigma_{2}^{j} \sigma_{3}^{k}
                           f_{abc} \lambda_{1}^{a} \lambda_{2}^{b} \lambda_{3}^{c}
                 \right] \delta^{(3)}(\vec{r}_{1}-\vec{r}_{2}) \delta^{(3)}(\vec{r}_{2}-\vec{r}_{3}),
 \label{eq:III3}
\end{eqnarray*}
with a coupling constant $V_{0}^{(3)}$ and the delta functions as a point-like three body interaction.
We can deduce the two body instanton induced force,
\begin{eqnarray*}
 H_{I\!I\!I}^{(2)} &=& \frac{V_{0}^{(2)}}{2} \sum_{i<j}
                                \left[ 1 + \frac{3}{32} \vec{\lambda}_{i} \!\cdot\! \vec{\lambda}_{j}
							 + \frac{9}{32} \vec{\sigma}_{i} \!\cdot\! \vec{\sigma}_{j}
							 					               \vec{\lambda}_{i} \!\cdot\! \vec{\lambda}_{j}
                               \right]  \delta^{(3)}(\vec{r}_{i}-\vec{r}_{j}),
   \label{eq:III2}
\end{eqnarray*}
using the quark condensate $\langle \bar{\psi}\psi \rangle$, where the coupling constant $V_{0}^{(2)}$ is given as
\begin{eqnarray*}
 V_{0}^{(2)} = \frac{1}{2} \langle \bar{\psi}\psi \rangle V_{0}^{(3)}.
\end{eqnarray*}
The interactions in the $q_{1}\bar{q}_{2}\bar{q}_{3}$ state are also obtained in a straightforward way.

\begin{figure}[tbp]
\begin{center}
\includegraphics[width=7cm, angle=0, clip]{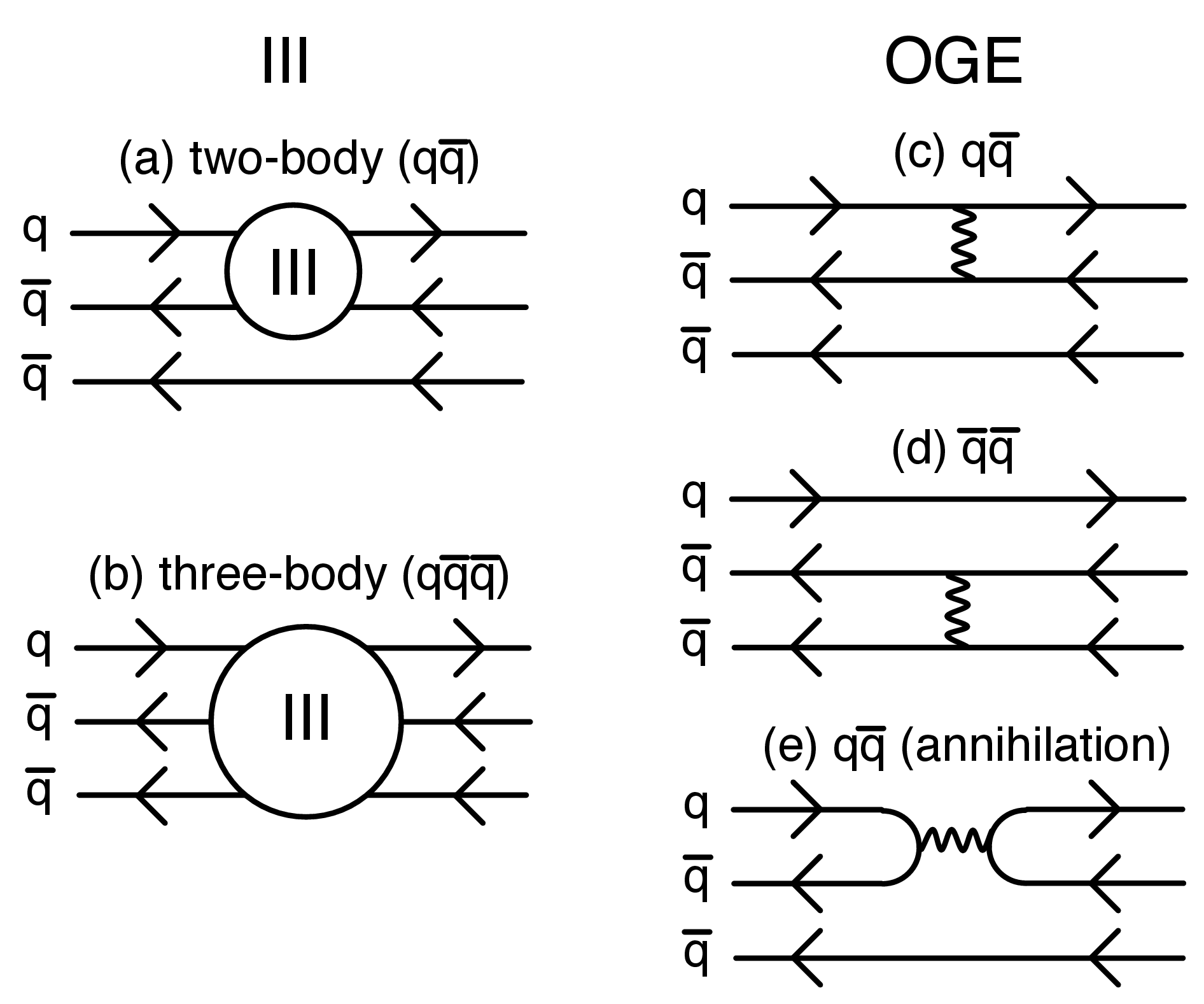}
\end{center}
\vspace*{0.0cm} \caption{\small \baselineskip=0.5cm The diagram contributions for the III and OGE potentials. III: (a) the two-body interaction for $q\bar{q}$ and (b) the three body interaction for $q\bar{q}\bar{q}$. OGE: (c) $q\bar{q}$, (d) $\bar{q}\bar{q}$ and (e) $q\bar{q}$ (annihilation).}
 \label{fig:diagram}
\end{figure}

The OGE potential between the $q_{1}q_{2}$ pair is given as
\begin{eqnarray*}
  V_{OGE} = 4 \pi \alpha_{S} \frac{\vec{\lambda}_{1} \!\cdot\! \vec{\lambda}_{2}}{4}
   \left[ \frac{1}{q^{2}} - \frac{\vec{\sigma}_{1} \!\cdot\! \vec{\sigma}_{2}}{6m_{1}m_{2}} \right],
  \label{eq:OGE}
\end{eqnarray*}
with a coupling constant $\alpha_{S}$.
The first term is the electric interaction, and the second the magnetic interaction with spin dependence.
However, we neglect the electric interaction, since in general it is sufficiently small as compared with the magnetic interaction.
It should be noted that the magnetic interaction is switched off with a suppression of $1/m_{Q}$ for the heavy-light quark pair (Qq) in the limit of the heavy mass.
Therefore, it is understood that the triquark $q\bar{q}\bar{q}$ may exist as a compound unit in the tetraquark structure.
As a summary, our interaction is sketched in Fig.~\ref{fig:diagram}.
The parameter set in our interaction \cite{Oka_Takeuchi_91} is summarized in Table~\ref{tbl:parameter}.

\begin{table}[tdp]
\begin{center}
\begin{tabular}{|c|c|c|c|c|}
\hline
$V_{0}^{(2)}$ &   $-\langle \bar{\psi}\psi \rangle$ &           $m_{q}$ &   $\omega$ & $\alpha_{s}$ \\
\hline
 -0.2564  [GeV$\,\mbox{fm}^{3}$] & $(0.25)^{3}$ [$\mbox{GeV}^{3}]$  & $0.3837$ [GeV] & $0.5$ [GeV]  & 1.319 \\
\hline                      
\end{tabular}
\end{center}
\label{default}
\caption{The parameter set from \cite{Oka_Takeuchi_91}.}
\label{tbl:parameter}
\end{table}%

From the III and OGE potentials, the energy spectrum of the triquark is obtained in the following way.
By using the basis of the spin and color, $\{ | \lambda X \rangle, | \rho Y \rangle, | \rho X \rangle, | \lambda Y \rangle \}$, we obtain the hamiltonian in matrix forms for flavor ${\overline{\bf 3}}_{A}$, ${\overline{\bf 3}}_{S}$, $\bf 6$ and $\overline{\bf 15}$ representations, respectively.

First we consider the III potential.
For the flavor ${\overline{\bf 3}}_{A}$ and ${\overline{\bf 3}}_{S}$ states, the hamiltonian is given in the basis $\{ | \lambda X \rangle, | \rho Y \rangle, | \rho X \rangle, | \lambda Y \rangle \}$ by
\begin{eqnarray}
H_{III}({\overline{\bf 3}}) = 
\left(
\begin{array}{ccccc}
               \frac{3}{2} &                           0 & \frac{15\sqrt{6}}{8}  &  -\frac{21}{4} \\
                             0 &                           3 &              -\frac{3}{4} &  -\frac{3\sqrt{6}}{4}  \\
 \frac{15\sqrt{6}}{8} &            -\frac{3}{4}  &           -\frac{21}{8}  &  \frac{9\sqrt{6}}{8}   \\
 -\frac{21}{4}           & -\frac{3\sqrt{6}}{4}  &   \frac{9\sqrt{6}}{8}  &  -\frac{15}{4} 
\end{array}
\right)
V_{0}^{(2)} I_{2}
+
\left(
\begin{array}{ccccc}
            -\frac{27}{4} &  \frac{9\sqrt{6}}{4} & 0 & 0 \\
    \frac{9\sqrt{6}}{4} &            -\frac{9}{2} & 0 & 0 \\
                             0 &                           0 & 0 & 0  \\
                             0 &                          0  & 0 & 0 
\end{array}
\right)
V_{0}^{(3)} I_{3},
\end{eqnarray}
where $I_{2}$ and $I_{3}$ are the expectation values of the delta function for the point-like interaction,
\begin{eqnarray}
I_{2} = \langle \Psi | \delta^{(3)}({\bf r}_{1}-{\bf r}_{2}) | \Psi \rangle = \left( \frac{m_{q}\omega}{2\pi} \right)^{3/2}
\end{eqnarray}
for the two-body interaction, and
\begin{eqnarray}
I_{3} = \langle \Psi | \delta^{(3)}({\bf r}_{1}-{\bf r}_{2}) \delta^{(3)}({\bf r}_{2}-{\bf r}_{3}) | \Psi \rangle
= \left( \frac{m_{q}\omega}{\sqrt{3}\pi} \right)^{3}
\end{eqnarray}
for the three-body interaction with the triquark spatial wavefunction $\Psi$.
It should be mentioned that the ${\overline{\bf 3}}_{A}$ and ${\overline{\bf 3}}_{S}$ states are mixed due to the off-diagonal element in the two-body interaction in the III potential, since $\{ | \lambda X \rangle, | \rho Y \rangle \}$ belongs to $\overline{\bf 3}_{A}$ and $\{ | \rho X \rangle, | \lambda Y \rangle \}$ to $\overline{\bf 3}_{S}$.
Therefore, we may denote the mixed state as ${\overline{\bf 3}}$ in the following discussion.
In the similar way, for the flavor $\bf 6$ state, the basis  $\{ | \lambda X \rangle, | \rho Y \rangle \}$ gives the matrix
\begin{eqnarray}
H_{III}({\bf 6}) =
\left(
\begin{array}{cc}
 \frac{63}{8} & -\frac{9\sqrt{6}}{8} \\
 -\frac{9\sqrt{6}}{8}  & \frac{21}{4}  
\end{array}
\right)
V_{0}^{(2)} I_{2}
+
\left(
\begin{array}{cc}
 \frac{27}{4} & -\frac{9\sqrt{6}}{4} \\
 -\frac{9\sqrt{6}}{4}  & \frac{9}{2}  
\end{array}
\right)
V_{0}^{(3)} I_{3},
\end{eqnarray}
and for the flavor $\bf \overline{\bf15}$ state, the basis $\{ | \rho X \rangle, | \lambda Y \rangle \}$ gives
\begin{eqnarray}
H_{III}({\overline{\bf 15}}) =
\left(
\begin{array}{cc}
 \frac{21}{8} & -\frac{9\sqrt{6}}{8} \\
 -\frac{9\sqrt{6}}{8}  & \frac{15}{4}  
\end{array}
\right)
V_{0}^{(2)} I_{2}.
\end{eqnarray}

Second we consider the OGE potential.
For the $\overline{\bf 3}$ state, we obtain the matrix in the basis of $\{ | \lambda X \rangle, | \rho Y \rangle, | \rho X \rangle, | \lambda Y \rangle \}$,
\begin{eqnarray}
V_{OGE}(\overline{\bf 3}) =
\left(
\begin{array}{ccccc}
            -\frac{7}{12} & \frac{5\sqrt{6}}{36} & \frac{\sqrt{6}}{36}  &  -\frac{1}{18} \\
 \frac{5\sqrt{6}}{36} &              -\frac{1}{6} &              -\frac{1}{6} &  \frac{\sqrt{6}}{18}  \\
   \frac{\sqrt{6}}{36} &            -\frac{1}{6}  &              \frac{1}{3}  &  \frac{\sqrt{6}}{9}   \\
 -\frac{1}{18}           &   \frac{\sqrt{6}}{18}  &   \frac{\sqrt{6}}{9}  &  -\frac{1}{18} 
\end{array}
\right)
\frac{4\pi \alpha_{s}}{m_{q}^{2}} I_{2}.
\end{eqnarray}
Furthermore, for the $\bf 6$ state, we obtain in the basis  $\{ | \lambda X \rangle, | \rho Y \rangle \}$ 
\begin{eqnarray}
V_{OGE}({\bf 6}) =
\left(
\begin{array}{cc}
      -\frac{11}{18} & \frac{\sqrt{6}}{6} \\
 \frac{\sqrt{6}}{6}  & -\frac{1}{3}  
\end{array}
\right)
\frac{4\pi \alpha_{s}}{m_{q}^{2}} I_{2},
\end{eqnarray}
and for the $\overline{\bf 15}$ state
\begin{eqnarray}
V_{OGE}(\overline{\bf 15}) =
\left(
\begin{array}{cc}
           \frac{1}{6} & \frac{\sqrt{6}}{6} \\
 \frac{\sqrt{6}}{6}  & -\frac{1}{9}  
\end{array}
\right)
\frac{4\pi \alpha_{s}}{m_{q}^{2}} I_{2},
\end{eqnarray}
in the basis  $\{ | \rho X \rangle, | \lambda Y \rangle \}$.

The mass differences among $u$, $d$ and $s$ quarks induce mixings between the flavor representations.
In the basis of the flavor representation, $\{ | {\overline{\bf 3}}_{A} \rangle,  | {\overline{\bf 3}}_{S} \rangle,  | {\overline{\bf 15}}^{0} \rangle, | {\bf 6} \rangle, | {\overline{\bf 15}}^{1} \rangle \}$, we easily obtain the mass part of the hamiltonian for $S=+1$ sector, as
\begin{eqnarray}
M_{mass} = 
\left(
\begin{array}{ccc|cc}
m_{u}+m_{d}+m_{s} &                                               0 &                                               0 & m_{u}-m_{d} &0   \\
                              0 & \frac{m_{u}+m_{d}}{2}+2m_{s} &  -\frac{m_{u}+m_{d}}{2}+m_{s}  & 0 & \frac{m_{u}-m_{d}}{\sqrt{2}} \\
   				       0 & -\frac{m_{u}+m_{d}}{2}+m_{s}  &                                                0 & 0 & -\frac{m_{u}-m_{d}}{\sqrt{2}}\\
\cline{1-5}
           m_{u}-m_{d} &                                                0 &                                                0  & m_{u}+m_{d}+m_{s} & 0\\
                             0 &       \frac{m_{u}-m_{d}}{\sqrt{2}} &       -\frac{m_{u}-m_{d}}{\sqrt{2}} &0 & 0 
\end{array}
\right).
\label{eq:mass}
\end{eqnarray}
The diagonal elements are isosinglet and isotriplet components, while the off-diagonal elements induce mixings between them.
Note that the flavor representations with the same symmetry ($A$ or $S$) are mixed.
The ${\overline{\bf 3}}_{A}$ and ${\overline{\bf 15}}^{0}$ states are mixed with each other by the $SU(3)$ symmetry breaking ( $m_{u}=m_{d}<m_{s}$).
We also note that the isosinglet states (${\overline{\bf 3}}_{A}, {\overline{\bf 3}}_{S}, {\overline{\bf 15}}^{0}$) and the isotriplet states (${\bf 6}, {\overline{\bf 15}}^{1}$) are also mixed due to the isospin symmetry breaking ($m_{u}<m_{d}$)\footnote{In the works in \cite{Dmitra_04, Dmitra_05, Dmitra_06}, the ${\overline{\bf 3}}_{A}$ and $\bf 6$ states are mixed due to the $m_{u}=m_{d} \neq m_{s}$.
As long as the isospin symmetry is not violated, however, we have no mixing between the ${\overline{\bf 3}}_{A}$ and $\bf 6$ states.}.
We consider this interaction as a driving force for the isospin symmetry breaking in the next section.
It should be noted that the Coulomb or electromagnetic interaction may also break isospin symmetry, which is not considered in this study.

\section{Isospin mixing}

In general, the $u-d$ quark mass difference is sufficiently small as compared with the energy splitting between the isosinglet and isotriplet states, and the isospin breaking can be neglected.
However, in the triquark, 
we see that the isosinglet and isotriplet states sometimes happen to be degenerate and thus a large isospin mixing can occur.
In this section, we investigate the mixing of the isosinglet and isotriplet states.
For this purpose, we calculate the eigenenergies, $E$, of the hamiltonian (\ref{eq:hamiltonian2}).
We choose the $s$ quark mass $m_{s}=0.48$ GeV and the strength of the harmonic potential $\omega=0.50$ GeV in the following discussion.
We take the parameter $p_{III}$ as a free parameter.

\begin{figure}[tbp]
\begin{center}
\includegraphics[width=7cm, angle=0, clip]{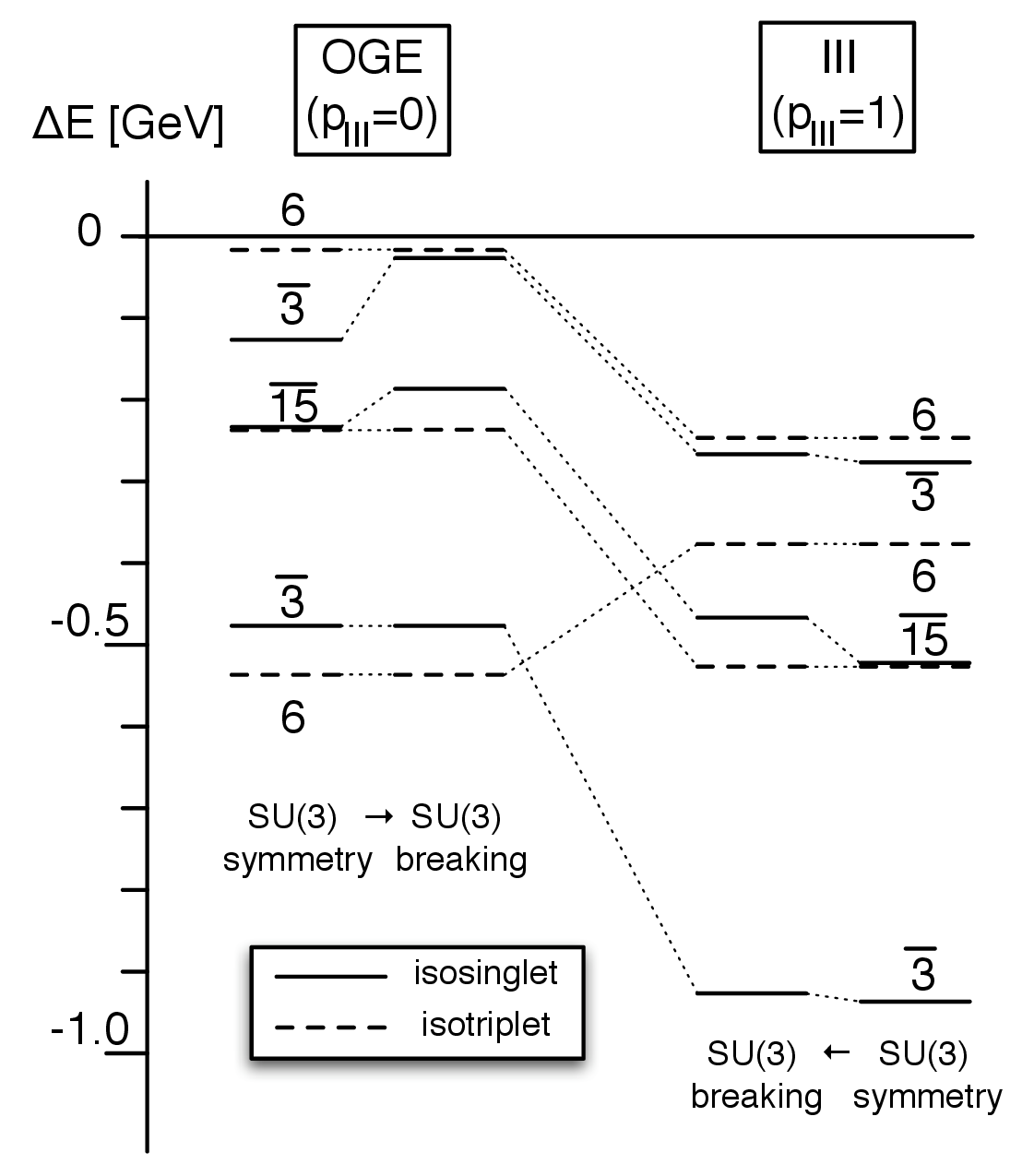}
\end{center}
\vspace*{0.0cm} \caption{\small \baselineskip=0.5cm The binding energies of the triquarks with $J=1/2$ for the OGE ($p_{III}=0$) and III ($p_{III}=1$) potentials.}
 \label{fig:spectrum}
\end{figure}

We present the binding energy spectrum of the triquark, $\Delta E = E - (m_{u}+m_{d}+m_{s})$, for the OGE ($p_{III}=0$) and III ($p_{III}=1$) potentials in Fig.~\ref{fig:spectrum}.
The isosinglet and isotriplet states are shown by the solid and dashed lines, respectively.
As the $J=3/2$ states are heavier than  $J=1/2$, in the following discussion, we pay attention to the ground states with the spin $J\!=\!1/2$, the $\overline{\bf 3}$ and $\bf 6$ multiplets.

Let us see the result by the III potential.
In $SU(3)$ symmetric case, the ground state is the $\overline{\bf 3}$ state, which contains mainly the ${\overline{\bf 3}}_{A}$ component rather than the ${\overline{\bf 3}}_{S}$ component.
On the other hand, the ${\overline{\bf 3}}_{S}$ component is mixed in the excited state in the $\overline{\bf 3}$ state.
Now we break the $SU(3)$ symmetry with keeping the isospin symmetry; $m_{u}=m_{d}<m_{s}$.
The ground state is still the $\overline{\bf 3}$ state, and the first excited state is the $\overline{\bf15}^{1}$ state, followed by the $\overline{\bf 15}^{0}$ and $\bf 6$ states.
The splitting between the $\overline{\bf 15}^{0}$ and $\overline{\bf 15}^{1}$ states makes the former lifted up as compared with the latter.
This splitting comes from the fact that the mass matrix (\ref{eq:mass}) mixes the ${\overline{\bf 3}}_{S}$ and $\overline{\bf 15}^{0}$ states.
The same mixing pushes the $\overline{\bf 3}$ upward.

On the other hand, in the OGE potential, the ground state is the $\bf 6$ state, followed by the $\overline{\bf 3}$, $\overline{\bf15}^{1}$, and $\overline{\bf15}^{0}$ states.
It should be noticed that the flavor multiplets are different in the III and the OGE potentials.
Especially the change of the ground state flavor is important for the isospin symmetry breaking as we see below.

The reason that the $\bf 6$ state is the ground state in the OGE can be understood by examining the annihilation diagram in Fig.~\ref{fig:diagram}(e).
It vanishes for the usual color singlet meson $q\bar{q}$, since the gluon ($g$) contained in the process $q\bar{q} \rightarrow g \rightarrow q\bar{q}$ is a color octet state.
In the triquark, however, the annihilation diagram does not vanish.
This is because the $g\bar{q}$ state contained in the process $q\bar{q}\bar{q} \rightarrow g\bar{q} \rightarrow q\bar{q}\bar{q}$ remains color anti-triplet ${\overline{\bf 3}}^{c}$ due to the color decomposition,
\begin{eqnarray*}
      {\bf 8}^{c} \otimes \overline{\bf 3}^{c} = \overline{\bf 3}^{c} \oplus {\bf 6}^{c} \oplus \overline{\bf 15}^{c}.
\end{eqnarray*}
Note that the initial and final $q\bar{q}\bar{q}$ states are also color anti-triplet.
The annihilation term increases the energy of the flavor $\overline{\bf 3}$ state, while it does not operate for the $\bf 6$ state (see Eq. (\ref{eq:wavefunction}) ).
Consequently, the $\overline{\bf 3}$ state is about 50 MeV above the $\bf 6$ state in the OGE potential.

\begin{figure}[tbp]
\begin{center}
\includegraphics[width=7cm, angle=0, clip]{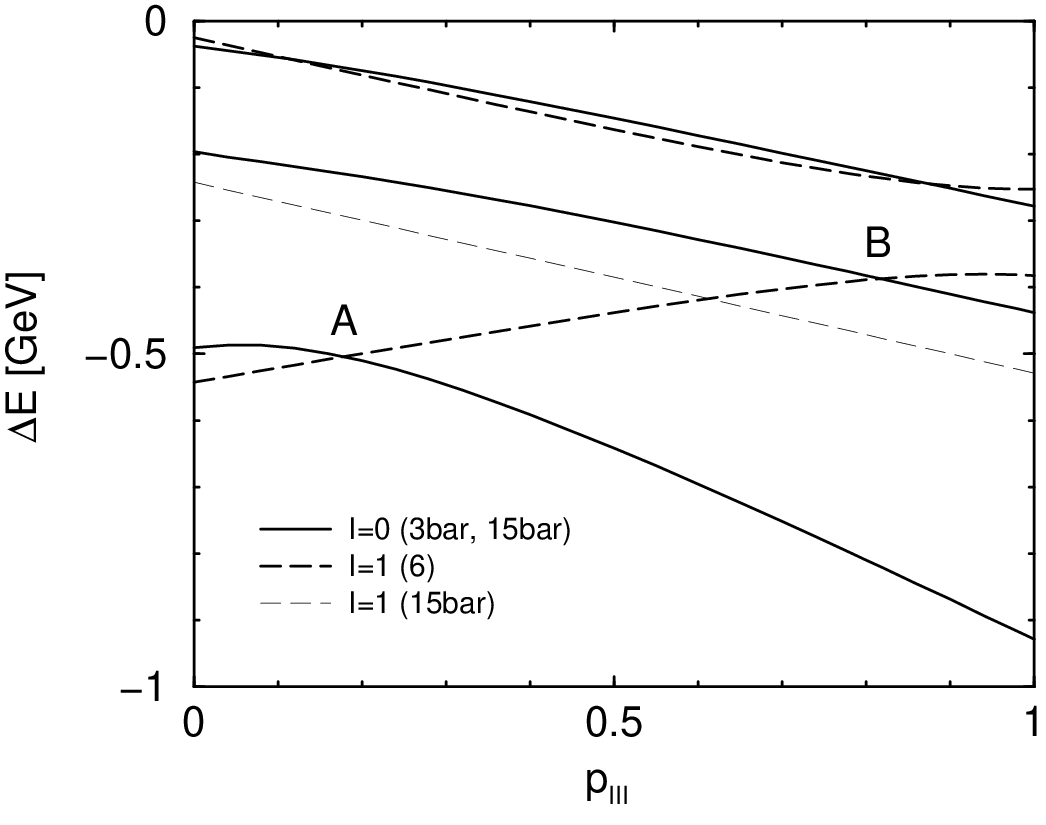}
\end{center}
\vspace*{0.0cm} \caption{\small \baselineskip=0.5cm The binding energies of the various flavor multiplets with $SU(3)$ breaking as functions of the parameter $p_{III}$. The bold-solid line indicates the isosinglet (${\overline{\bf 3}}$ and ${\overline{\bf 15}}^{0}$) states, the bold-dashed line isotriplet (${\bf 6}$), and the thin-dashed line the isotriplet (${\overline{\bf 15}}^{1}$) states. Cf. Fig.~\ref{fig:spectrum}.}
 \label{fig:E_I0_I1_pIII}
\end{figure}

Let us return to the discussion of the isospin symmetry breaking.
We recall that the $\overline{\bf 3}$ state is isosinglet and the $\bf 6$ state is isotriplet.
Thus the mixing of the flavor multiplets are directly related to the isospin mixing.
Explicitly, we plot the binding energies of the flavor multiplets as functions of the parameter $p_{III}$ in Fig.~\ref{fig:E_I0_I1_pIII}.
The solid lines indicate the isosinglet states, and the dashed lines the isotriplet states.
We find that the isosinglet and isotriplet states become degenerate at A ($p_{III}=0.18$) and B ($p_{III}=0.82$).

\begin{figure}[tbp]
\begin{minipage}{8cm}
\vspace*{0.0cm}
\centering
\includegraphics[width=7cm]{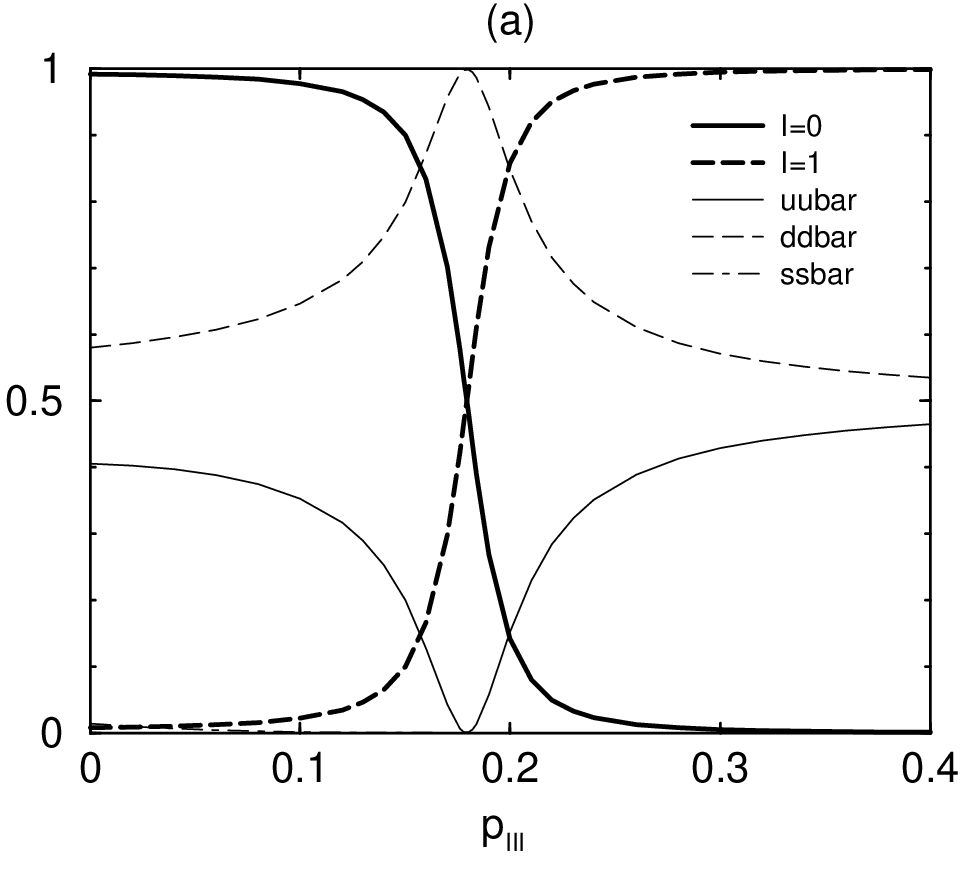}
\vspace{-0.0cm}
\end{minipage}
\begin{minipage}{8cm}
\centering
\includegraphics[width=7cm]{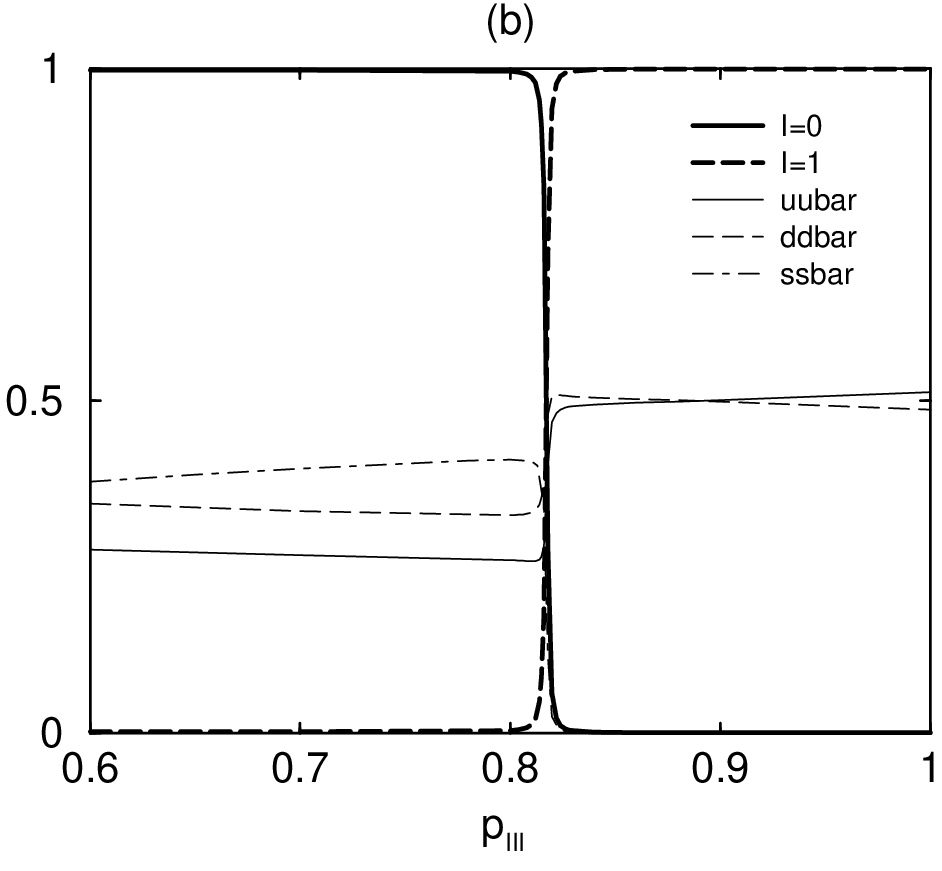}
\vspace{-0cm}
\end{minipage}
\caption{\small \baselineskip=0.5cm The ratios of the isosinglet (bold-solid line) and isotriplet (bold-dashed line) states as functions of the parameter $p_{III}$. (a) and (b) corresponds to the state A and B, respectively, in Fig.~\ref{fig:E_I0_I1_pIII}. The ratios of the $u\bar{u}$ (thin-solid line),  $d\bar{d}$ (thin-dashed line) and  $s\bar{s}$ (thin-dot-dashed line) are also shown.}
    \label{fig:uu_dd_ss_pIII}
\end{figure}

Now let us introduce the isospin symmetry breaking, namely the $ud$ quark mass difference, $\Delta m = m_{d}-m_{u} \simeq 0.005$ GeV \cite{Nasu_etal_03}.
The $\Delta m$ is comparable to the energy difference between the isosinglet and isotriplet states at A and B.
There, the two degenerate states will split into two isospin mixed states which are orthogonal to each other.
Here we choose one state at A.
The ratios of isosinglet and isotriplet components are plotted as functions of the parameter $p_{III}$ in Fig.~\ref{fig:uu_dd_ss_pIII}(a).
In the range of $0.16 < p_{III} < 0.20$, we see a rapid change of the isosinglet (bold-solid line) and the isotriplet (bold-dashed line) components, hence the isospin is strongly mixed.
In the same way at B, we also see an isospin mixing at $p_{III}=0.82$ as shown in Fig.~\ref{fig:uu_dd_ss_pIII}(b).
However, in contrast to the case A, the isospin mixing at B occurs in a small range of the parameter $p_{III}$.
This is understood from the mass matrix (\ref{eq:mass}).
At A, the isosinglet state is almost the $\overline{\bf 3}_{A}$ multiplet, while the isotriplet state is purely the $\bf 6$ multiplet (see Fig.\ref{fig:spectrum}).
The mass matrix (\ref{eq:mass}) induces the $\overline{\bf 3}_{A}$ and $\bf 6$ multiplet mixing, namely the isospin violation, by $m_{u}-m_{d}$.
On the other hand, at B,  the isosinglet state is changed to be the $\overline{\bf 15}^{0}$ state, while the isotriplet state is the same.
In the mass matrix (\ref{eq:mass}), however, there is no direct mixing between the $\overline{\bf 15}^{0}$ and $\bf 6$ multiplets.
They are mixed indirectly through the multi-step mixings of the ${\bf 6} - \overline{\bf 3}_{A}$, $\overline{\bf 3}_{A} - \overline{\bf 3}_{S}$, and $\overline{\bf 3}_{S} - \overline{\bf 15}^{0}$.
Therefore the isospin mixing at B is suppressed as compared to that at A.

Here we recall the isospin violation in experimental observations.
Maiani {\it et al.}  consider $D_{s}(2632)$ as the $c\bar{s}d\bar{d}$ state, which is an isospin mixed state \cite{Maiani_etal_04}.
In our analysis, the isospin mixing at A induces a mixing between the isosinglet (mostly $\overline{\bf 3}_{A}$) and isotriplet ($\bf 6$) states.
Hence, from Eq.~(\ref{eq:wavefunction}), the mixed wavefunction, $| \overline{\bf 3}_{A} \rangle - | {\bf 6} \rangle = - d(\bar{d}\bar{s}-\bar{s}\bar{d})$, contains only the $d\bar{d}$ component.
On the other hand, at B, there is a mixing between the isosinglet (mostly $\overline{\bf 15}^{0}$) and the isotriplet ($\bf 6$) states.
There, from Eq.~(\ref{eq:wavefunction}), the mixed wavefunction $| \overline{\bf 15}^{0} \rangle - | {\bf 6} \rangle$ contains both of the $u\bar{u}$ and $d\bar{d}$ components with the same fraction.
Consequently, we see that the isospin mixed states, $c\bar{s}u\bar{u}$ and $c\bar{s}d\bar{d}$, become separate eigenstates by the $\overline{\bf 3}-{\bf 6}$ mixing rather than the $\overline{\bf 15}^{0}-{\bf 6}$ mixing.

We also understand this result explicitly by looking at the fraction of $u\bar{u}$, $d\bar{d}$ and $s\bar{s}$ components at A and B in Fig.~\ref{fig:uu_dd_ss_pIII}(a) and (b), respectively.
In Fig.~\ref{fig:uu_dd_ss_pIII}(a), the $d\bar{d}$ fraction (thin-dashed line) is overwhelming as compared with the $u\bar{u}$ fraction (thin-solid line) around $p_{III}=0.18$.
In contrast, in Fig.~\ref{fig:uu_dd_ss_pIII}(b), the fraction of the  $u\bar{u}$ and $d\bar{d}$ components are almost the same at $p_{III}=0.82$.
Therefore, the isospin mixed state at $A$ gives the $c\bar{s}d\bar{d}$, while the state at B does not.
Thus, the discussion by Maiani {\it et al.}  in \cite{Maiani_etal_04} is proven to be possible as the the $\overline{\bf 3}-{\bf 6}$ mixing.

So far, we have discussed the isospin mixing by using the isospin basis of isosinglet and isotriplet.
However, the isospin mixing is also investigated by basis $\{ u\bar{u}, d\bar{d} \}$.
Then the hamiltonian is generally given by
\begin{eqnarray}
\bordermatrix{
              & u\bar{u} & d\bar{d} \cr
u\bar{u} & m & \delta \cr
d\bar{d} & \delta &m+2\Delta m \cr
}.
 \label{eq:ud_matrix_1}
\end{eqnarray}
The $u\bar{u}$ and $d\bar{d}$ are eigenstates of this hamiltonian, if the flavor mixing term $\delta$ is much smaller than $\Delta m$, and only the diagonal component is dominant.
However, in general, $\delta$ is in the order of hundred MeV in the vacuum as we see the mass splitting of $\pi-\eta$.

\begin{figure}[tbp]
\begin{minipage}{8cm}
\vspace*{0.0cm}
\centering
\includegraphics[width=7cm]{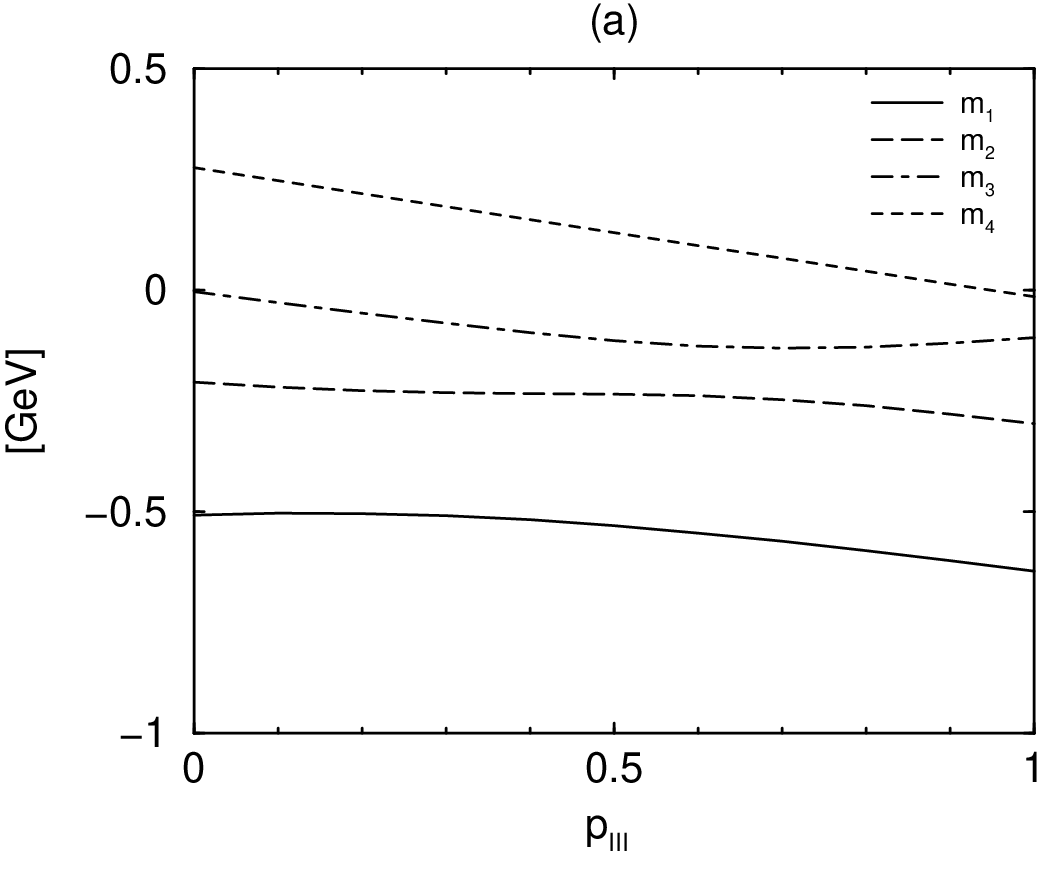}
\vspace{-0.0cm}
\end{minipage}
\begin{minipage}{8cm}
\centering
\includegraphics[width=7cm]{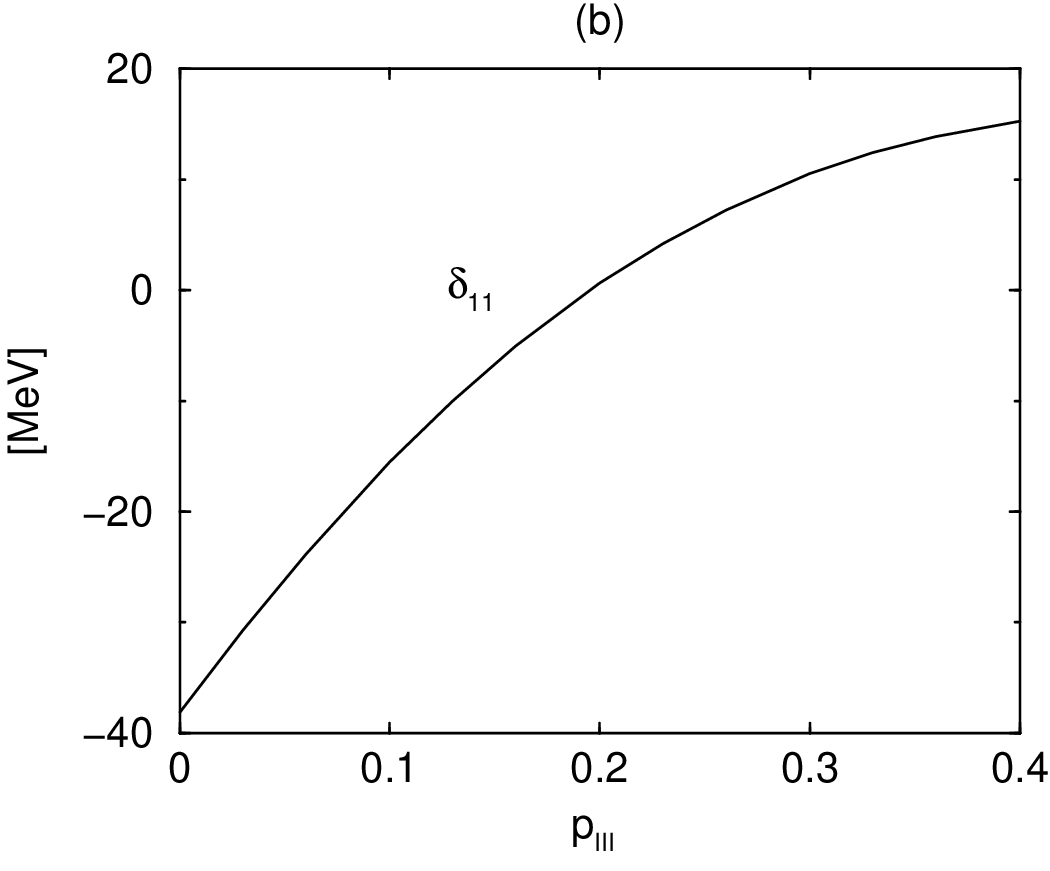}
\vspace{-0cm}
\end{minipage}
\caption{\small \baselineskip=0.5cm (a) The diagonal components of the $u\bar{u}-d\bar{d}$ matrix as functions of the parameter $p_{III}$. (b) $\delta_{11}$ as a function of $p_{III}$. Note the energy unit is given by MeV in (b). See the text.} 
    \label{fig:ud_uu_pIII_w50}
\end{figure}

For the triquark with $J=1/2$, due to the combination of spin and color, we have four $u\bar{u}$-like states and also four $d\bar{d}$-like states.
In this basis, the hamiltonian is given by
\begin{eqnarray}
\begin{array}{r@{ }l}
& \begin{array}{cccccc}
\mbox{ } & \makebox[2.0em]{ } & u\bar{u} & \makebox[12em]{ } & d\bar{d} &
\end{array} \\
\begin{array}{l}
   \raisebox{4ex} {$u\bar{u}$} \\ \raisebox{-4ex} {$d\bar{d}$}
 \end{array}
&\left(
\begin{array}{cccc|cccc}
 m_{1} &   &   & 0  & \delta_{11} & \delta_{12} & \delta_{13} & \delta_{14} \\
     & m_{2} &  &  & \delta_{21} & \delta_{22} & \delta_{23} & \delta_{24}   \\
     &   & m_{3} & &  \delta_{31} & \delta_{32} & \delta_{33} & \delta_{34}   \\
0 &  & & m_{4}  &  \delta_{41} & \delta_{42} & \delta_{43} & \delta_{44}   \\
\cline{1-8}
\delta_{11} & \delta_{12} & \delta_{13} & \delta_{14}  &  m_{1} \!+\! 2\Delta m &   &   & 0 \\
\delta_{21} & \delta_{22} & \delta_{23} & \delta_{24}  &   & m_{2} \!+\! 2\Delta m  &   &  \\
\delta_{31} & \delta_{32} & \delta_{33} & \delta_{34}  &   &   & m_{3} \!+\! 2\Delta m  &  \\
\delta_{41} & \delta_{42} & \delta_{43} & \delta_{44}  & 0 &   &   &  m_{4} \!+\! 2\Delta m
\end{array}
\right),
\end{array}
 \label{eq:ud_matrix_2}
\end{eqnarray}
where the diagonal $u\bar{u}-u\bar{u}$ and $d\bar{d}-d\bar{d}$ parts are diagonalized in the spin and color  spaces.
The diagonalized energy, $m_{1}$, $m_{2}$, $m_{3}$ and $m_{4}$, are plotted as functions of the parameter $p_{III}$ in Fig.~\ref{fig:ud_uu_pIII_w50}(a).
If the flavor mixing strength $\delta_{ij}$ ($i,j=1,\cdots, 4$) are sufficiently small, the lowest $u\bar{u}$- and $d\bar{d}$-like states become eigenstates.
As shown in Fig.~\ref{fig:ud_uu_pIII_w50}(b), $\delta_{11}$ is so small as compared with $\Delta m$ around $p_{III}=0.18$.
There, the eigenstate become $u\bar{u}$- and $d\bar{d}$-like states, hence the isosinglet and isotriplets states are ideally mixed.
This result is consistent with our discussion that the isospin mixing is caused by the $\overline{\bf 3}-{\bf 6}$ mixing around $p_{III}=0.18$.
It should be noted that the contribution from the higher states is suppressed since the mixing is in the order of $\delta_{ij}/(m_{k}-m_{1}) \simeq 0.1$ ($k \ge 2$) in the perturbation theory.
Therefore the first order perturbation is sufficient for the present discussion.

Lastly, we discuss the parameter dependence of the isospin mixing.
We employ several free parameters, the $s$ quark mass $m_{s}$, the harmonic oscillator potential frequency $\omega$ and the parameter $p_{III}$ for the OGE and III potentials.
They may have some uncertainty due to the lack of the experimental information.
However, one sees that the results are not modified qualitatively by parameter change.
As an example, we plot the size parameter $b=1/\sqrt{m_{q}\omega}$ of the triquark wavefunction, which causes the $\overline{\bf 3}-{\bf 6}$ mixing at A, 
as a function of the parameter $p_{III}$ for $m_{s}=0.48$ GeV and $0.58$ GeV.
We see that the $b$ comes within a reasonable range $0.4 < b < 0.6$ fm, and is not far from $b=0.5$ fm \cite{Oka_Takeuchi_91}.
This range is little affected by $m_{s}$.

Concerning the range of the parameter $p_{III}$, the obtained value $p_{III}=0.18$ is smaller than the conventionally used value $p_{III} \simeq 0.4$ in hadron spectroscopy \cite{Oka_Takeuchi_91}.
This observation indicates that the OGE is more dominant than the instanton induced interaction in tetraquarks.
It is noticed that the value $p_{III}=0.18$ is not obtained dynamically, since the quark wave function is assumed to be Gaussian.
The present study suggests that there would exist an essential mechanism to choose such $p_{III}$ in charmed tetraquark.

When the linear potential is used as a confinement potential, the quark wave function is modified from that of the harmonic oscillator potential, and the absolute values of the OGE and the instanton induced interaction are also modified.
However, the ratio of both interactions is not changed, since both of the potentials are point-like interactions.
In the present discussion, the isospin symmetry breaking is induced by the ratio of two interactions.
Therefore our conclusion is not modified qualitatively.

\begin{figure}[tbp]
\begin{center}
\includegraphics[width=7cm, angle=0, clip]{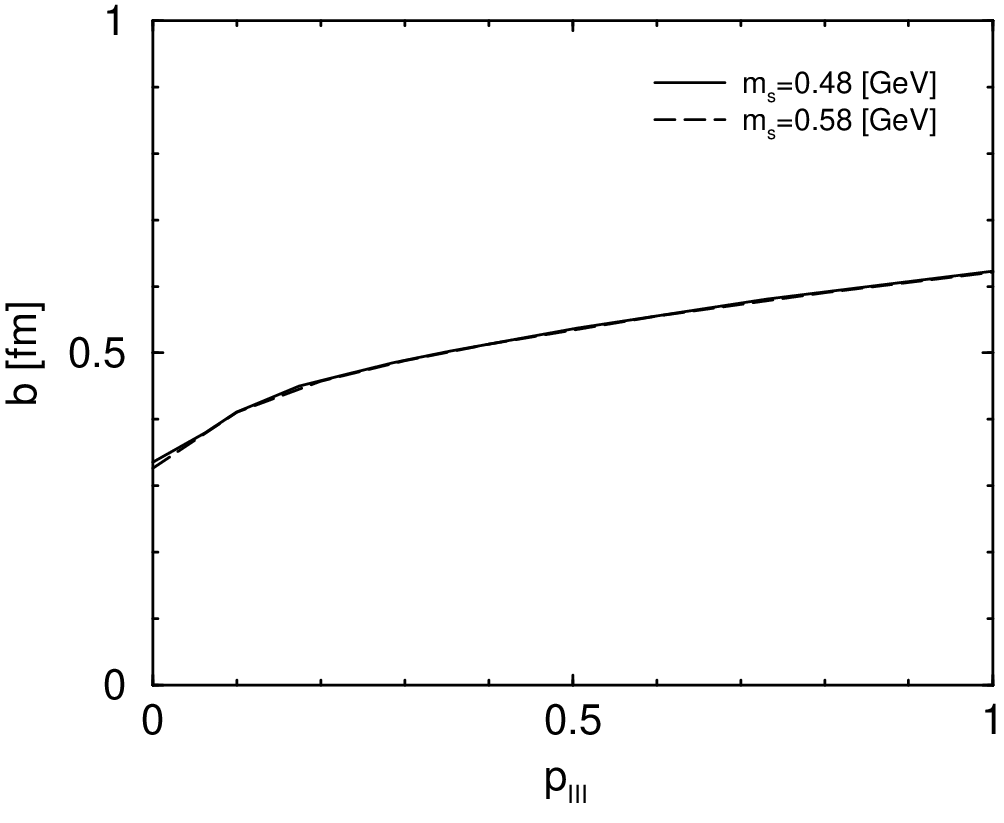}
\end{center}
\vspace*{0.0cm} \caption{\small \baselineskip=0.5cm The $b-p_{III}$ relation for the $\overline{\bf 3}-{\bf 6}$ mixing. $m_{s}=0.48$ GeV (solid line) and $0.58$ GeV (dashed line). See the text.}
 \label{fig:b_w_pIII_ms48_58}
\end{figure}

\section{Conclusion}

Possibility of isospin violation in the $D_{s}$ tetraquark systems is examined in this paper.
Tetraquarks are candidates of the exotic $D_{s}$ mesons recently reported in experiment.
We consider the energy spectrum of the triquark by using the non-relativistic quark model with the instanton induced interaction and the one-gluon exchange potentials.
With taking the $SU(3)$ symmetry breaking into account for $S=+1$ sector, 
we show that the flavor $\overline{\bf 3}$ (isosinglet) and the flavor $\bf 6$ (isotriplet) representations form the ground states.
Considering the isospin symmetry breaking by the quark mass difference, $m_{u}<m_{d}$, it is shown that the $\overline{\bf 3}$ (isosinglet) and ${\bf 6}$ (isotriplet) states may be mixed strongly with some range of the parameter $p_{III}$.
There the isosinglet and the isotriplet states are ideally mixed, 
and one of the eigenstates is dominated by the $d\bar{d}$ component.
This result is also investigated by looking at the off-diagonal components in the $u\bar{u}-d\bar{d}$ matrix.
Our conclusion supports the discussion given in \cite{Maiani_etal_04, Maiani_etal_05}.

How do we experimentally confirm the picture given in this paper?
The present mechanism of the isospin symmetry violation relies on the suppression of the flavor mixing interaction.
Thus, at the ideal (maximal) mixing, the $u\bar u$- and $d\bar d$-like states are split by the diagonal part of the mass matrix, namely by $2\Delta m \sim 10$ MeV.
Therefore the two states are expected to come close to each other.
So far, due to the experimental restriction, only a few charged decay modes are observed, and they suggest a $d\bar d$-like state, $D_s^+ (2630/ c\bar s d\bar d) $, where its main decay mode is
$D_s^+(2630) \to D_s\eta$, while $D_s^+(2630) \to D^0 K^+$ is suppressed.
The corresponding $u\bar u$-like state will show different decay patterns.
Therefore careful analyses of different charged modes of decays will reveal the nature of the isospin breaking.  In particular, the decays into $D_s \pi^0$ and $D^+ K^0$ are two interesting modes.

The present study suggests the possibility of the triquark  \`a la  ``color non-singlet baryon", which is a color non-singlet particle composed by three quarks.
Although the triquark itself cannot exist asymptotically, it may appear as an effective degree of freedom in the exotic heavy mesons in heavy quark mass limit.
It is considered in general that the color non-singlet light quark systems may exist by color neutralization with heavy quark spectator \cite{Jaffe_05}.
The triquark is a possible candidate among the color non-singlet quark systems, which can be examined by studying the tetraquark structure of exotic open charm mesons.
The triquark would be also an interesting object in the lattice QCD simulation.
Furthermore the triquark may be a relevant degree of freedom as a color non-singlet compound particle in the deconfinement phase such as the quark-gluon plasma and the quark matter.
In order to understand such states in many aspects, it is important to study several properties, such as masses, decay widths and so forth. 

\section*{Acknowledgment}
We express our thanks to Dr.~T.~Shinozaki and Prof.~S.~Takeuchi for discussions.
This work is supported by a Grant-in-Aid for Scientific Research for Priority Areas, MEXT (Ministry of Education, Culture, Sports, Science and Technology) with No. 17070002.


\end{document}